\documentclass{aa}

\usepackage{graphicx}
\usepackage{amsmath}
\usepackage{txfonts}
\usepackage{booktabs}
\usepackage{threeparttable}
\usepackage{xcolor}
\usepackage{hyperref}
\usepackage{CJKutf8}
\usepackage{booktabs,array,tabularx}
\usepackage{placeins}
\usepackage{float}

\graphicspath{{figures/}}

\newcommand{\cii}{[\ion{C}{ii}]}
\newcommand{\lcii}{L_{\rm \cii}}
\newcommand{\ltir}{L_{\rm TIR}}
\newcommand{\lfir}{L_{\rm FIR}}
\newcommand{\msun}{M_\odot}
\newcommand{\lsun}{L_\odot}
\newcommand{\kms}{\mathrm{km\,s^{-1}}}
\newcommand{\angstrom}{\mbox{\AA}}
\newcommand{\yssl}{J1440$+$0019}

\begin{document}

\title{A dust-reddened quasar in a \cii-luminous host at $z=6.55$}

\author{Daming Yang (\begin{CJK*}{UTF8}{gbsn}羊达明\end{CJK*})\inst{\ref{aff1}}\thanks{E-mail: \texttt{dyang@strw.leidenuniv.nl}}
\and Joseph F.~Hennawi\inst{\ref{aff1},\ref{aff2}}
\and Rychard Bouwens\inst{\ref{aff1}}
\and Timo Kist\inst{\ref{aff1}}
\and Eduardo Ba\~nados\inst{\ref{aff4}}
\and \\
Hiddo S.~B.~Algera\inst{\ref{aff8}}
\and Aaron J.~Barth\inst{\ref{aff5}}
\and Sarah E.~I.~Bosman\inst{\ref{aff3},\ref{aff4}}
\and Frederick B.~Davies\inst{\ref{aff4}}
\and Anna-Christina Eilers\inst{\ref{aff9},\ref{aff10}}
\and \\
Xiaohui Fan\inst{\ref{aff7}}
\and Victoria A.~Fawcett\inst{\ref{aff15},\ref{aff16}}
\and Yuming Fu\inst{\ref{aff1}}
\and Anniek J.~Gloudemans\inst{\ref{aff13}}
\and Francesco Guarneri\inst{\ref{aff11}}
\and Bin Jia\inst{\ref{aff1}}
\and \\Elia Pizzati\inst{\ref{aff1},\ref{aff14}}\thanks{NHFP Einstein Fellow}
\and Huub R\"ottgering\inst{\ref{aff1}}
\and Jan-Torge Schindler\inst{\ref{aff11}}
\and Feige Wang\inst{\ref{aff12}}
\and Jinyi Yang\inst{\ref{aff12}}
}

\institute{Leiden Observatory, Leiden University, P.O. Box 9513, 2300 RA Leiden, The Netherlands\label{aff1}
\and Department of Physics, University of California, Santa Barbara, CA 93106, USA\label{aff2}
\and Institute for Theoretical Physics, Heidelberg University, Philosophenweg 12, D-69120 Heidelberg, Germany\label{aff3}
\and Max-Planck-Institut f\"ur Astronomie, K\"onigstuhl 17, 69117 Heidelberg, Germany\label{aff4}
\and Department of Physics and Astronomy, 4129 Frederick Reines Hall, University of California, Irvine, CA 92697-4575, USA\label{aff5}
\and Steward Observatory, University of Arizona, 933 North Cherry Avenue, Tucson, AZ 85718, USA\label{aff7}
\and NSF NOIRLab, Gemini Observatory, 670 N A'ohoku Place Hilo, HI 96720, USA\label{aff13}
\and Institute of Astronomy and Astrophysics, Academia Sinica, 11F of Astronomy-Mathematics Building, No.\,1, Sec.\ 4, Roosevelt Rd, Taipei 106319, Taiwan, R.O.C.\label{aff8}
\and MIT Kavli Institute for Astrophysics and Space Research, 77 Massachusetts Avenue, Cambridge, MA 02139, USA\label{aff9}
\and Department of Physics, Massachusetts Institute of Technology, 77 Massachusetts Avenue, Cambridge, MA 02139, USA\label{aff10}
\and Hamburg Observatory, University of Hamburg, Gojenbergsweg 112, D-21029 Hamburg, Germany\label{aff11}
\and European Southern Observatory (ESO), Karl-Schwarzschild-Strasse 2, 85748 Garching bei M\"unchen, Germany\label{aff15}
\and School of Mathematics, Statistics and Physics, Newcastle University, NE1 7RU, UK\label{aff16}
\and Center for Astrophysics $\vert$ Harvard \& Smithsonian, Cambridge, MA 02138, USA\label{aff14}
\and Department of Astronomy, University of Michigan, 1085 South University Avenue, Ann Arbor, MI 48109, USA\label{aff12}
}

\date{Received date, accepted date}

\abstract{Most known quasars at \(z>6\) were selected through their rest-frame UV emission. Only a small number of reddened quasars have been identified at these redshifts, leaving their nuclear and host-galaxy properties poorly constrained. We present JWST/NIRSpec spectroscopy and ALMA band~6 observations of the quasar \yssl\ at \(z=6.5488\), revealing its dust-reddened nature and jointly characterizing its nucleus and host galaxy. The continuum is best described by an SMC-like extinction law with \(E(B-V)=0.20\pm0.03\). Three absorption systems detected in \ion{N}{v}, \ion{Si}{iv}, and \ion{C}{iv} are blueshifted by approximately \(900\)--\(3600~\kms\) and likely trace an ionized nuclear outflow. The broad \ion{Mg}{ii} line and extinction-corrected continuum yield a black-hole mass of \(\log(M_{\rm BH}/\msun)=9.04\pm0.42\), a bolometric luminosity of \(L_{\rm bol}\simeq1.1\times10^{47}~{\rm erg\,s^{-1}}\), and an Eddington ratio of \(\lambda_{\rm Edd}\simeq0.8\). ALMA observations further reveal its host galaxy to be among the most \cii-luminous known at \(z>6\), with \(\lcii=(7.21\pm0.25)\times10^9~\lsun\). For a fiducial dust temperature of 47 K and emissivity index \(\beta=1.6\), the dust continuum implies \(\ltir=8.9\times10^{12}~\lsun\) and \(M_{\rm dust}=4.0\times10^8~\msun\). The inferred dynamical mass places \yssl\ in the region of the \(M_{\rm BH}\)--\(M_{\rm dyn}\) plane occupied by high-redshift blue quasars. These observations reveal a moderately reddened, rapidly accreting quasar in a luminous, dust-rich host during reionization. Its measured nuclear and host-galaxy properties provide an observational reference for identifying and characterizing similar systems in future systematic searches with wide-area infrared surveys.}


\maketitle
\nolinenumbers

\section{Introduction}
\label{sec:introduction}

Optical and near-infrared surveys have discovered several hundred quasars during reionization (\(z\gtrsim6\)) so far, nearly all of them selected through their bright rest-frame UV emission (e.g., \citealp{Fan2001,Jiang2016,Banados2016,Matsuoka2016}; J.~\citealp{Yang2023}; D.~\citealp{Yang2026}). This selection is strongly biased against quasars with even modest dust reddening. For an SMC-like extinction law, \(E(B-V)\gtrsim0.3\) can cause even intrinsically luminous quasars (\(M_{1450}\sim-28\)) to be missed by deep optical searches such as Subaru High-z Exploration of Low-Luminosity Quasars \citep[SHELLQs;][]{Kato2020}. 
X-ray studies suggest that the obscured AGN fraction increases from low to high redshift \citep[reviewed by][]{Hickox2018}, with \(\sim60\)--\(80\%\) of AGN at \(3\leq z<6\) obscured by column densities of \(N_{\rm H}>10^{23}\,\mathrm{cm^{-2}}\) \citep{Vito2018}. 
A related trend is seen in comparisons of eROSITA X-ray and optical/UV luminosity functions, which suggest that the fraction of luminous AGN (\(L_{2-10\,\mathrm{keV}}\gtrsim10^{45}\,\mathrm{erg\,s^{-1}}\)) missed by UV selection increases with redshift, reaching \(\gtrsim80\%\) at \(z\sim4\)--6 \citep{Roster2026}. Comparisons of JWST/MIRI-selected AGN luminosity functions with those of UV-selected quasars infer even higher UV-obscured fractions of \(\sim98\)--\(99\%\) for luminous AGN at \(4.5<z<6\) \citep{Bulichi2026}.
While these estimates depend on selection and modeling assumptions, complementary evidence suggests that UV-bright quasar phases trace only part of early black-hole growth.
Constraints on the ionizing output and lifetimes of high-redshift quasars suggest that their observed UV-bright phases cannot readily account for the assembly of their massive black holes \citep{Davies2019,Durovcikova2025,Durovcikova2026}.
Combining quasar clustering with abundance measurements at \(z\sim6\)--7 likewise favors low UV-bright duty cycles, with central estimates of order one percent or less \citep{Eilers2024,Pizzati2024,Huang2026}.
Together, these findings strengthen the scenario for a substantial contribution from obscured accretion to early black-hole growth. However, only a small number of reddened quasars have been identified at \(z\gtrsim6\), leaving both their abundance and physical properties poorly constrained.


The limited sample available so far has been assembled through a variety of selection methods. Among optically selected SHELLQs quasars, \citet{Kato2020} used optical-to-mid-infrared SED fitting to identify two moderately reddened (\(E(B-V)\simeq0.12\)) broad-line objects at \(z>5.6\), including J1205$-$0000 \citep{Izumi2021}. Near-infrared photometry has since identified additional red-quasar candidates within the SHELLQs sample \citep{Iwamoto2025}.
JWST follow-up of SHELLQs sources with luminous narrow Ly\(\alpha\) emission and weak UV continua has revealed broad Balmer emission in several objects, with Balmer decrements indicating BLR extinctions of \(A_V\sim1\)--2 \citep{Matsuoka2025}. In deep-field observations, GNz7q at \(z=7.19\) was identified as a compact red UV source using HST imaging and grism spectroscopy, with an initial reddening estimate of \(E(B-V)\simeq0.11\) \citep{Fujimoto2022}. Its quasar nature was subsequently confirmed through broad Balmer emission in JWST spectra \citep{Fei2026}. COS-87259 at \(z=6.85\) was initially selected as a Lyman-break galaxy candidate, before its infrared and radio emission revealed a heavily obscured, radio-loud AGN \citep{Endsley2022}. SED modeling implies a rest-frame \(9.7\,\mu{\rm m}\) optical depth of \(\tau_{9.7}\simeq2.3\) \citep{Endsley2023}. 
JWST/MIRI imaging from the Systematic Mid-infrared Instrument Legacy Extragalactic Survey \citep[SMILES;][]{Lyu2024} and the MIRI Early Obscured AGN Wide Survey \citep[MEOW;][]{Bulichi2026,Leung2026} is also uncovering obscured AGN and candidates through their hot-dust emission, including candidates at \(z>6\). The narrow-line AGN identified in MEOW reside in diverse host environments, with evidence for obscuration on both circumnuclear and host-galaxy scales \citep{Leung2026}. These discoveries span from moderately reddened broad-line quasars to heavily obscured sources, but how these systems are related remains uncertain. Detailed observations of both the nucleus and the host galaxy are needed to investigate the origin of the reddening and its connection to the host's gas and dust.

In this paper, we present JWST and ALMA observations of HSC J144045.91$+$001912.9 (\(\alpha=220.19129^{\circ}\), \(\delta=+0.32025^{\circ}\), J2000; hereafter \yssl), a quasar at \(z=6.55\) discovered in the SHELLQs survey \citep{Matsuoka2022}. 
The source was selected through its Lyman break. Its published \(K\)-band magnitude is \(K_{\rm AB}=21.72\pm0.32\) \citep{Matsuoka2022}. We also identify a match in CatWISE2020 \citep{Marocco2021} with \(W1_{\rm AB}=20.56\pm0.11\) and \(W2_{\rm AB}=20.31\pm0.15\), suggesting a red rest-frame UV-to-optical continuum and possible reddening.
Our JWST/NIRSpec spectroscopy further reveals a dust-reddened continuum, while ALMA observations detect bright \cii\ and dust continuum emission from the host galaxy. Together, these data allow us to characterize the nuclear reddening and black-hole accretion properties alongside the cold gas and dust in the host, and to compare the blueshifted absorption seen toward the nucleus with the host-galaxy gas kinematics.

We describe the observations and data reduction in Sect.~\ref{sec:data} and present the measurements in Sect.~\ref{sec:results}. In Sect.~\ref{sec:discussion} we discuss the location of the obscuring dust, the far-infrared and \cii\ properties of the host, the position of \yssl\ in the \(M_{\rm BH}-M_{\rm dyn}\) plane, the possible connection between the blueshifted absorption and the host gas, and the relation of \yssl\ to red quasar populations at low and high redshift. We summarize our conclusions in Sect.~\ref{sec:conclusions}. Throughout this paper we adopt the Planck 2018 cosmology \citep{Planck2020}; at \(z=6.5488\), one arcsecond corresponds to 5.55 kpc, and the cosmic microwave background temperature is \(T_{\rm CMB}\simeq20.6\) K.

\section{Observations and data reduction}
\label{sec:data}

\subsection{JWST/NIRSpec spectroscopy}
\label{sec:jwst}

\yssl\ was observed with the JWST/NIRSpec fixed slit as part of the cycle 4 GO program 9180 (PI: J.~Hennawi), which obtained grating spectroscopy of 27 faint quasars at \(6.5<z<7.7\) in order to measure the reionization history of the intergalactic medium from their Ly\(\alpha\) damping wings. Twelve of the program targets were discovered by \textit{Euclid} \citep{Yang2026}, while the remainder, including \yssl, are from SHELLQs \citep[e.g.,][]{Matsuoka2016,Matsuoka2022}. 
The inclusion of \yssl\ in the JWST program was motivated primarily by its strong \cii\ detection in earlier ALMA observations as described in Sect.~\ref{sec:alma}.
The observations used the G140H/F070LP and G235H/F170LP grating--filter combinations, providing a nominal resolving power of \(R\simeq2700\) \citep{Jakobsen2022}. Together, they cover observed wavelengths of 0.812--3.127 \(\mu{\rm m}\), corresponding to rest-frame wavelengths of 1075--4140 \(\angstrom\) for \yssl. The exposure times were 5894\,s in G140H and 4143\,s in G235H, split equally between the S200A1 and S200A2 slits to cover the gap between the two detectors. For G140H/F070LP, we extract beyond the standard red limit of 1.27 \(\mu{\rm m}\), set by second-order contamination. This is possible because the contaminating light originates at half the observed wavelength, which for \yssl\ lies blueward of Ly\(\alpha\), where IGM absorption leaves effectively zero flux.

We reduced the data with a combination of the standard JWST calibration pipeline and a custom pipeline built on \texttt{PypeIt} \citep{Prochaska2020}. The JWST pipeline (version 1.20.2, with CRDS context \texttt{jwst\_1464.pmap}) performed the detector-level and two-dimensional spectroscopic calibration, and \texttt{PypeIt} carried out the sky subtraction, optimal extraction, and coaddition of the individual exposures. We refer the reader to \citet{Yang2026euvlg} and Yang et al. (2026, in prep.) for a full description of the procedure.


We rescaled the coadded spectrum by a factor of 1.396 to match the HSC \(y\)-band photometry. This correction may partly reflect residual slit losses, although intrinsic variability between the HSC and JWST epochs could also contribute. Figure~\ref{fig:photometry} compares the spectrum before this rescaling with the available HSC, VIKING, and CatWISE2020 photometry. The \(K_s\)-band flux lies below the synthetic photometry derived from the unscaled JWST spectrum, and the discrepancy increases after the \(y\)-band rescaling. The origin of this mismatch remains uncertain and requires further investigation, including a reassessment of the archival \(K_s\)-band photometry. We use the \(y\)-band-scaled spectrum for the following analysis.


\subsection{ALMA observations}
\label{sec:alma}
\yssl\ was observed with ALMA in the cycle-10 program CISTERN
\citep[\cii-SelecTed Early universe RefereNce; project 2023.1.00443.S, PI:
R.~Bouwens][]{Bouwens2026}, which targets the \cii\ 158~\(\mu{\rm m}\) line in 92 UV-bright sources with spectroscopic redshifts. The observations were executed on 2024 March 7 with 45 antennas of the 12-m array in the compact C-1 configuration, with an on-source time of 5.5~min. The band 6 correlator setup consisted of four 1.875 GHz spectral windows centered at 249.27, 251.39, 263.28, and 266.40 GHz, covering the redshifted \cii\ line (in the 251.39 GHz window) and the adjacent continuum. Further details of the survey design and source-by-source results will be presented in R.~J.~Bouwens et al.~2026 (in prep.).

The measurement set was restored from the QA2 delivery products with the
observatory-supplied \texttt{scriptForPI.py} script in CASA
\citep{CASAteam2022}. The visibilities were time-averaged into 30~s intervals
and imaged with the \textsc{tclean} task using natural weighting. At the line frequency the synthesized beam is \(1.39\times1.16\)
arcsec at a position angle of \(73^\circ\), the native channel width is
\(9.3~\kms\), and the channel rms is \(0.67~{\rm mJy\,beam^{-1}}\).

The continuum image was formed from the line-free 248.40--250.15 GHz spectral
window; its effective frequency is 249.276 GHz and its map rms is
\(0.17~{\rm mJy\,beam^{-1}}\).
The line cube was continuum subtracted before the moment maps were
constructed.
We measured the integrated line spectrum with a curve-of-growth analysis
centered on the continuum source, and for the final luminosity we adopt the
flux within an elliptical aperture matched to the beam
(\(1.39\times1.16\) arcsec). This choice depends less on a parametric spatial
model than does the somewhat larger flux obtained by fitting a marginally
resolved disk. The systemic redshift of \yssl, \(z=6.5488\), is measured
from this line (Sect.~\ref{sec:cii}).

\section{Results and analysis}
\label{sec:results}

\subsection{A dust-reddened quasar}
\label{sec:reddening}
 
\begin{figure*}
  \centering
  \includegraphics[width=\textwidth]{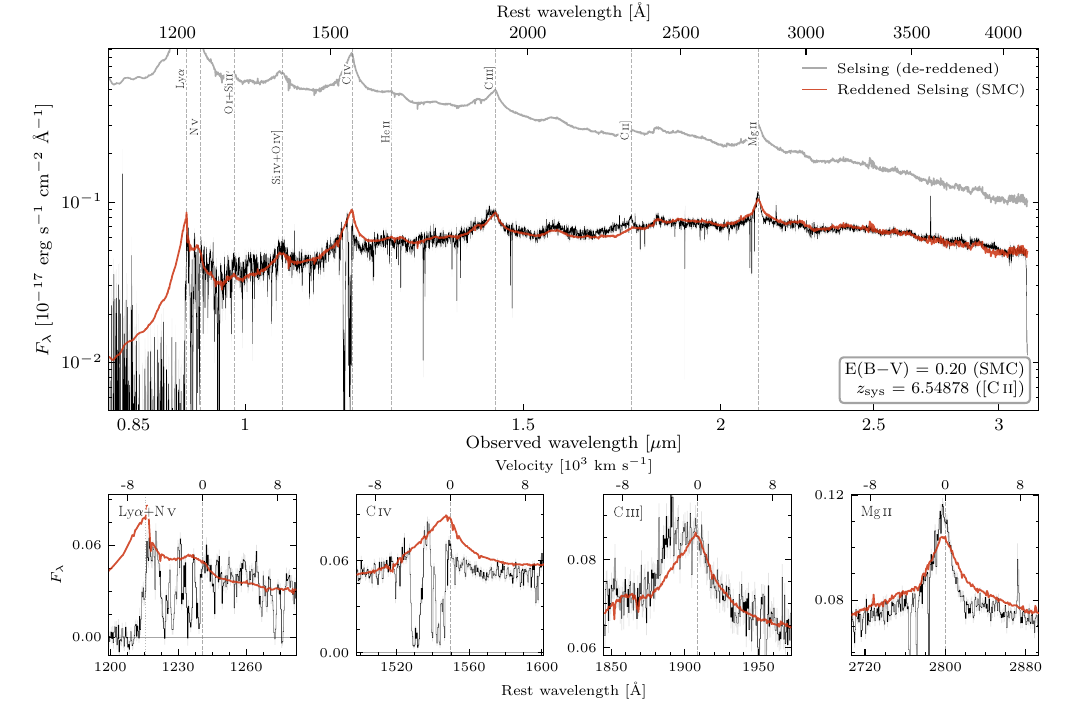}
  \caption{JWST/NIRSpec spectrum of \yssl\ (\(z_{\rm sys}=6.5488\)). \emph{Top}: the black line shows the G140H+G235H spectrum. The gray curve shows the blue type-1 quasar composite of \citet{Selsing2016}, and the red curve shows the same composite attenuated by the best-fit SMC law with \(E(B-V)=0.20\) (Sect.~\ref{sec:reddening}). Vertical dashed lines mark typical quasar broad emission lines. \emph{Bottom}: zoom-ins of the Ly\(\alpha\)+\ion{N}{v}, \ion{C}{iv}, \ion{C}{iii}], and \ion{Mg}{ii} regions. Blueshifted absorption is present in \ion{N}{v} and \ion{C}{iv}, as well as in \ion{Si}{iv} (Sect.~\ref{sec:absorption}, Fig.~\ref{fig:absorbers}). Narrow \ion{Mg}{ii} absorption is also detected.}
  \label{fig:overview}
\end{figure*}

Figure~\ref{fig:overview} shows the NIRSpec spectrum in black. The continuum is clearly reddened compared with the quasar composite of \citet{Selsing2016} (gray). At the same time, several broad emission lines are detected, including \ion{Si}{iv} \(\lambda\)1394, \ion{C}{iv} \(\lambda\)1548, \ion{C}{iii}] \(\lambda\)1909, and \ion{Mg}{ii} \(\lambda\)2796. This combination identifies \yssl\ as a red type-1 quasar: the nucleus is viewed directly, but through moderate dust extinction along the sightline. Several high-ionization transitions show blueshifted absorption, and narrow absorption features appear around \ion{Mg}{ii} (see Sect.~\ref{sec:absorption} and Figs.~\ref{fig:overview} and \ref{fig:absorbers}).
 
To inspect the reddening, we fit the observed continuum with the blue quasar composite of \citet{Selsing2016} attenuated by an extinction law. The free parameters are the color excess \(E(B-V)\) and the normalization. We adopt the SMC extinction curve of \citet{Gordon2003} with \(R_V=2.74\), the common choice for reddened type-1 quasars, whose continua are in general well described by steep, SMC-like extinction without the 2175 \(\angstrom\) feature \citep[e.g.,][]{Richards2003,Hopkins2004,Glikman2012}. The fit is evaluated in line-free rest-frame windows between 1285 and 3090 \(\angstrom\)\footnote{1285--1295, 1315--1330, 1350--1360, 1445--1465, 1690--1705, 1770--1810, 2190--2230, 2640--2670, and 3030--3090 \(\angstrom\).}, with the absorption systems of Sect.~\ref{sec:absorption} masked in addition. The uncertainty on \(E(B-V)\) is dominated by the uncertainty of the intrinsic spectrum rather than by statistical errors. We estimate this systematic uncertainty by changing the intrinsic continuum slope of the composite by \(\Delta\alpha_\lambda=\pm0.3\) and repeating the fit.

The SMC fit gives \(E(B-V)=0.195\pm0.03\). Changing the intrinsic slope by \(\pm0.3\) moves the value between 0.166 and 0.225, which sets the systematic uncertainty. A color excess of this amount places J1440+0019 at the mildly reddened end of the red type-1 population at lower redshift (Sect.~\ref{sec:lowz}).

For the fixed intrinsic template, the SMC extinction curve gives the lowest reduced \(\chi^2\), with \(\chi^2/\nu=3.6\), compared with 5.7--93.0 for the alternative extinction and attenuation laws tested in Appendix~\ref{app:extlaws}. We therefore quote the SMC value as a compact description of the observed reddening rather than as evidence for a particular dust composition.

We also note a weak emission feature near rest-frame \(2326~\angstrom\), consistent with \ion{C}{ii}]. The feature lies on a broader excess above the reddened composite around \(2300~\angstrom\). This excess may partly reflect differences in blended \ion{Fe}{ii} emission, which produces a broad shelf in this spectral region \citep{Vestergaard2001}.

\subsection{Absorption systems}
\label{sec:absorption}

\begin{figure}
  \centering
  \includegraphics[width=\hsize]{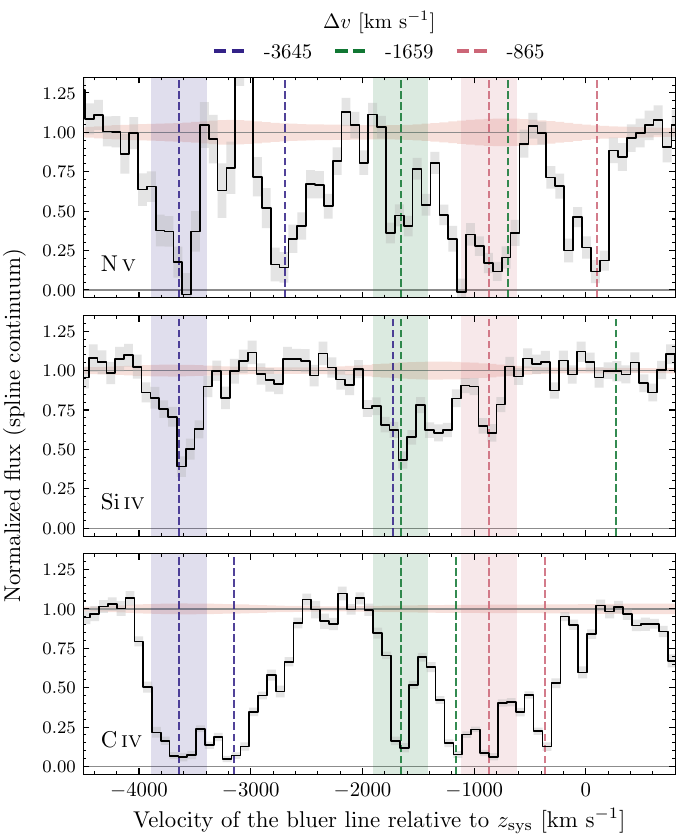}
  \caption{Continuum-normalized \ion{N}{v}, \ion{Si}{iv}, and \ion{C}{iv} profiles. Velocity is defined by the blue member of each doublet relative to \(z_{\rm \cii}=6.5488\).}
  \label{fig:absorbers}
\end{figure}

Figure~\ref{fig:overview} shows multiple absorption troughs blueward of the \ion{N}{v} \(\lambda\lambda\)1239, \ion{Si}{iv} \(\lambda\lambda\)1394, and \ion{C}{iv} \(\lambda\lambda\)1548 emission lines. In Fig.~\ref{fig:absorbers} we show the continuum-normalized\footnote{To obtain an empirical continuum model, we fitted a spline to line-free windows of the observed spectrum, masking the absorption troughs and the broad emission lines, and propagated the uncertainty in the continuum placement with Monte Carlo realizations of the fit, shown as the band around unity in Fig.~\ref{fig:absorbers}.} spectrum around these transitions. Three systems, at velocity offsets of \(-3645\), \(-1659\), and \(-865~\kms\) relative to the \cii\ redshift, are detected in all three ions. Creating these ions requires energies of 33.5 eV (\ion{Si}{iv}), 47.9 eV (\ion{C}{iv}), and 77.5 eV (\ion{N}{v}). In particular, \ion{N}{v} is rare in intervening gas and arises mostly in material exposed to a hard radiation field such as that of a quasar \citep{Fechner2009}. The presence of \ion{N}{v} and the corresponding absorption in multiple ions support an association with the quasar, with the blueshifts suggesting an outflow. These observations do not, however, constrain the distance of the absorbing gas from the nucleus. Several troughs approach zero normalized flux and are probably saturated, but the complex does not qualify as a broad absorption line (BAL), which requires a trough with a width over \(2000~\kms\) \citep{Weymann1991}: the deepest \ion{C}{iv} feature spans about \(1000~\kms\), which places the system in the mini-BAL regime \citep[\(500\)--\(2000~\kms\)][]{Hamann2004}. Because components of different systems overlap and the narrow cores are not resolved at \(R\simeq2700\), the doublet ratios cannot be measured reliably. We defer the detailed modeling of these systems, and with it any estimate of their covering fractions, column densities, and kinetic power, to future work. We also note that the other red type-1 quasar known at \(z>6\), the SHELLQs quasar J1205$-$0000 \citep{Matsuoka2016}, shows the same absorption signature in the same ions, there with troughs broad enough to meet the BAL criterion \citep{Onoue2019,Izumi2021,Sawamura2025}. We return to this comparison of high-\(z\) red sources in Sect.~\ref{sec:landscape}.

Strong low-ionization absorption is also present at about \(-3407~\kms\) (\(z\simeq6.463\)), as shown in Fig.~\ref{fig:lowion}. It is detected in \ion{Mg}{ii} \(\lambda\)2796, in \ion{Fe}{ii} \(\lambda\)2344, \(\lambda\)2383, \(\lambda\)2587, and \(\lambda\)2600, and in \ion{Si}{ii} \(\lambda\)1260, \ion{O}{i} \(\lambda\)1302, and \ion{C}{ii} \(\lambda\)1335. The origin of this low-ionization system remains uncertain. The absorption could arise in foreground gas or in gas associated with the quasar, including dense or shielded material in an outflow. Its different velocity from the high-ionization systems does not by itself distinguish these possibilities. The \ion{Mg}{ii} absorption lines are masked when fitting the broad \ion{Mg}{ii} emission (Sect.~\ref{sec:mbh}).

\subsection{\ion{Mg}{ii}, black-hole mass, and Eddington ratio}
\label{sec:mbh}

\begin{figure}
  \centering
  \includegraphics[width=\hsize]{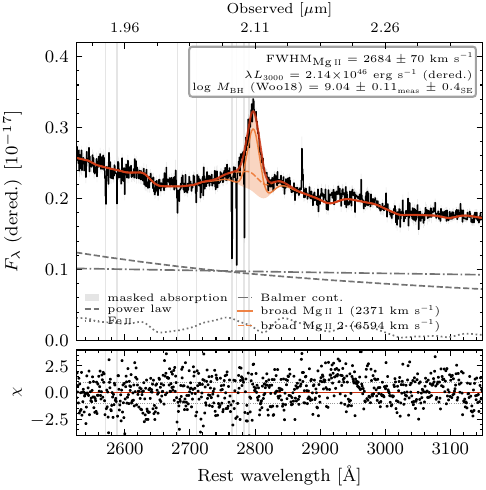}
  \caption{\ion{Mg}{ii} fit after correcting the spectrum for SMC-like reddening with \(E(B-V)=0.20\). The black curve shows the spectrum, and the orange curve shows the total best-fitting model. The orange shading marks the broad \ion{Mg}{ii} emission, modeled as the sum of two Gaussian components, above the fitted continuum. The dashed, dotted, and dash-dotted curves show the power-law, \ion{Fe}{ii}, and Balmer-continuum components, respectively. Gray bands mark narrow absorption excluded from the fit. The annotated black-hole-mass uncertainty separates the measurement and template term from the single-epoch calibration scatter. The lower panel shows residuals in units of the spectral uncertainty.}
  \label{fig:mgii}
\end{figure}

In this section we measure the broad \ion{Mg}{ii} line and the continuum luminosity at $\lambda_{\rm rest}=3000\,\angstrom$, and use them to estimate the black-hole mass ($M_{\rm BH}$) and Eddington ratio ($\lambda_{\rm Edd}$). Figure~\ref{fig:mgii} shows the \ion{Mg}{ii} region of the spectrum after correction for the \(E(B-V)=0.20\) SMC-like reddening of Sect.~\ref{sec:reddening}. We fitted the rest-frame 2700--2900 \(\angstrom\) region with a power-law continuum, the UV \ion{Fe}{ii} template of \citet{Tsuzuki2006}, a Balmer continuum, and two broad Gaussian components, with the continuum anchored in line-free windows on either side of the line and narrow absorption masked iteratively. The best-fit model is overplotted in Fig.~\ref{fig:mgii}, with the total model, the power-law continuum, and the combined \ion{Fe}{ii} plus Balmer-continuum component shown separately. The sum of the two broad components has \({\rm FWHM}_{\rm MgII}=2684\pm70~\kms\), and the fitted power law gives a dereddened continuum luminosity of \(\lambda L_\lambda(3000)=2.14\times10^{46}~{\rm erg\,s^{-1}}\), corresponding to \(7.9\times10^{45}~{\rm erg\,s^{-1}}\) before the reddening correction. The uncertainty is about 0.07 dex, accounting for the uncertain intrinsic continuum slope.

Using the \citet{Woo2018} relation, we obtain \(\log(M_{\rm BH}/\msun)=9.04\pm0.42\). The measurement and template term of 0.11 dex combines 0.03 dex from the line fit, 0.10 dex from the \ion{Fe}{ii} treatment, and 0.03 dex from the intrinsic-slope range propagated through the reddening correction. Adding the intrinsic scatter of the single-epoch calibration, about 0.4 dex, in quadrature gives a total uncertainty of 0.42 dex. Without the reddening correction the same fit yields \(\log(M_{\rm BH}/\msun)=8.83\), and a different attenuation law could shift the result further.

For the dereddened luminosity, the nominal Eddington ratio is \(\lambda_{\rm Edd}=L_{\rm bol}/L_{\rm Edd}\simeq0.8\). Because this number folds together the reddening correction, the bolometric correction, and the virial mass calibration, and because the 0.42 dex mass uncertainty alone maps it to a range of roughly 0.3 to 2.1, we quote it as a nominal value and do not take it as evidence for super-Eddington accretion. H\(\delta\) falls within the final 0.03 \(\mu{\rm m}\) of the useful G235H spectrum, but its position near the detector edge and the structure of the surrounding continuum do not support a stable independent virial fit, which leaves \ion{Mg}{ii} as the only virial estimator available here.

\subsection{\cii\ emission and dust continuum}
\label{sec:cii}

\begin{figure*}
  \centering
  \includegraphics[width=\textwidth]{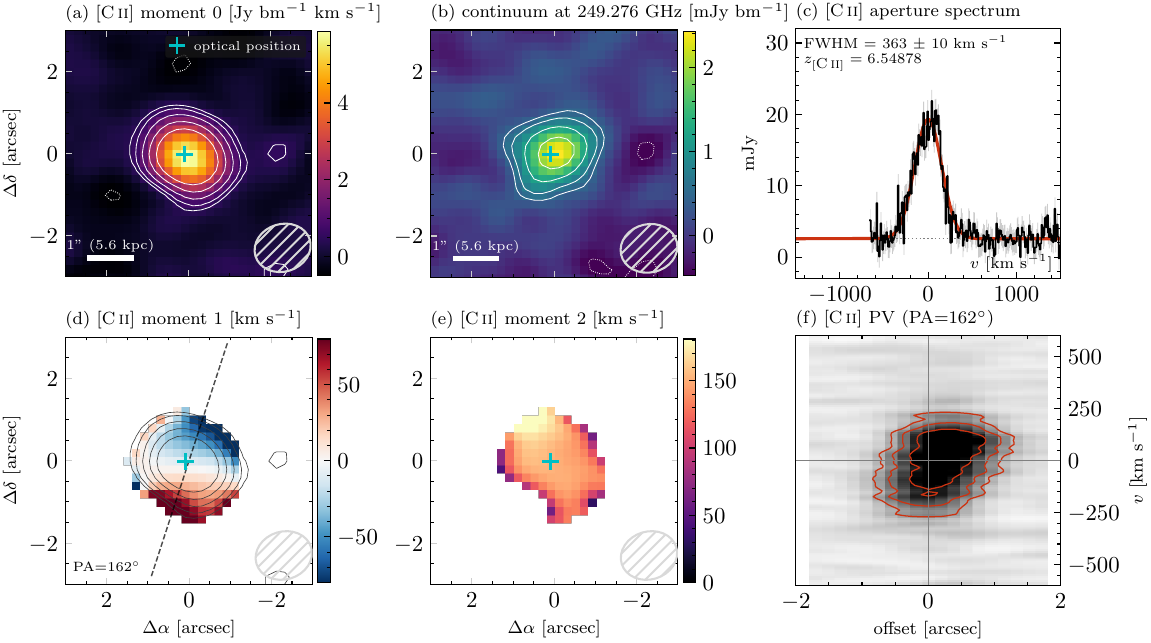}
  \caption{ALMA observations of \yssl. Panels a and b show the continuum-subtracted \cii\ moment-zero map integrated over \(-450<v<+450~\kms\) and the 249.3 GHz continuum map, respectively. White contours are \(-3\), \(-2\), 2, 3, 5, 8, and 12 times the map rms where they fall inside the plotted range. Panel c shows the \cii\ spectrum extracted within the beam-matched elliptical aperture described in Sect.~\ref{sec:cii}, with the single-Gaussian fit used to measure the systemic redshift and line FWHM shown in red. Panels d and e show the first- and second-moment maps within the connected source mask defined in Sect.~\ref{sec:cii}. Panel f shows the position--velocity diagram along the measured velocity gradient, extracted with a slit one synthesized-beam major-axis FWHM wide. Velocities are measured relative to \(z_{\rm \cii}=6.5488\). Crosses mark the optical position of the quasar, and hatched ellipses show the synthesized beam. The bright emission spans approximately one synthesized beam, limiting the kinematic interpretation.}
  \label{fig:alma}
\end{figure*}

Figure~\ref{fig:alma} summarizes the ALMA data for \yssl. Panel a shows the \cii\ moment-zero map of the continuum-subtracted cube integrated over \(-450<v<+450~\kms\), while panel b shows the 249.276 GHz continuum flux-density map formed by averaging the line-free channels from 248.40 to 250.15 GHz. Panel c shows the \cii\ spectrum extracted within an elliptical aperture with the size of \(1.39''\times1.16''\) and the same position angle as the synthesized beam.\footnote{The aperture size was determined from the curve of growth and encloses approximately 95 percent of the total \cii\ line flux.} The single-Gaussian fit used to determine the systemic redshift and line FWHM is shown in red. Panels d and e show the first- and second-moment maps evaluated from the unsmoothed continuum-subtracted cube within the connected source mask. The mask requires emission above \(3\sigma\) in at least two adjacent channels after spatial smoothing with a Gaussian whose FWHM is half the synthesized-beam major-axis FWHM. Panel f shows the position--velocity diagram along the measured velocity-gradient axis at \({\rm PA}=162^\circ\), extracted with a slit one synthesized-beam major-axis FWHM wide. Both the line and continuum are detected at high significance, and their centroids and the optical position of the quasar agree to within approximately 0.1 arcsec.

The spatially integrated continuum flux density is \(S_\nu=2.62\pm0.42~{\rm mJy}\) at an effective observed frequency of 249.276 GHz. The continuum map has a peak signal-to-noise ratio of 9.5. Describing the dust emission with a single modified blackbody at \(T_{\rm dust}=47\) K and \(\beta=1.6\), a common convention for high-redshift quasars \citep{Beelen2006}, including the cosmic microwave background correction of \citet{daCunha2013} and adopting \(\kappa_{850\mu{\rm m}}=0.77~{\rm cm^2\,g^{-1}}\) \citep{Dunne2000}, we obtain a far-infrared (FIR) luminosity of \(\lfir=6.3\times10^{12}~\lsun\), integrated over rest-frame 42.5--122.5 \(\mu{\rm m}\), a total infrared (TIR) luminosity of \(\ltir=8.9\times10^{12}~\lsun\), integrated over 8--1000 \(\mu{\rm m}\), and a dust mass of \(M_{\rm dust}=4.0\times10^8~\msun\). Converting all of \(\ltir\) to a star formation rate (SFR) with the \citet{Kennicutt2012} calibration gives \({\rm SFR_{IR}}\simeq1330~\msun\,{\rm yr^{-1}}\), which we treat as an upper limit because the quasar may heat part of the dust (Sect.~\ref{sec:fir}).

The \cii\ line gives \(z_{\rm \cii}=6.5488\pm0.0001\), \(S_{\rm \cii}\Delta v=6.47\pm0.23~{\rm Jy\,km\,s^{-1}}\), and \({\rm FWHM}_{\rm \cii}=363\pm10~\kms\), corresponding to \(L_{\rm \cii}=(7.21\pm0.25)\times10^9~\lsun\). The \cii\ emission is marginally resolved. A two-dimensional Gaussian fit to the moment-zero map gives a deconvolved FWHM size of \(0.69\times0.41\) arcsec, or \(3.8\times2.3\) kpc, but with a \(1.39\times1.16\) arcsec beam only the major axis is meaningfully constrained, and we use the size in Sect.~\ref{sec:fir} only to set a characteristic radius.

The host is bright in both \(L_{\rm \cii}\) and \(L_{\rm IR}\). It is among the most luminous \cii\ emitters known among \(z>6\) quasars \citep{Decarli2018, Venemans2020}, and its infrared luminosity places it at the upper end of the same samples. We compare J1440+0019 with other \cii-bright quasar hosts and discuss the dust-temperature dependence of the infrared quantities in Sect.~\ref{sec:fir}.

The first-moment map (Fig.~\ref{fig:alma}d) shows a velocity gradient from about \(-65\) to \(+65~\kms\) along a position angle near \(155^\circ\), consistent with the position--velocity diagram (Fig.~\ref{fig:alma}f). We discuss the kinematics and potential host structure in Sect.~\ref{sec:fir}.

The \cii\ profile appears slightly skewed, but the current data quality does not support an additional kinematic component, and no broad wing appears within the covered velocity range. None of the three mini-BAL systems described in Sect.~\ref{sec:absorption} (\(-865\), \(-1659\), and \(-3645~\kms\)) is covered by the current spectral setup. Whether the nuclear ionized outflow has a cold-gas counterpart at these velocities therefore cannot be tested with the present data and requires further observations. Table~\ref{tab:properties} collects these measurements together with the derived quantities of Sect.~\ref{sec:discussion}.

\begin{table*}
\caption{Measured and derived properties of \yssl.}
\label{tab:properties}
\centering
\begin{tabular}{lll}
\toprule
Quantity & Value & Comment \\
\midrule
\multicolumn{3}{l}{\emph{Measurements}} \\
\(z_{\rm \cii}\) & \(6.5488\pm0.0001\) & aperture spectrum \\
\(E(B-V)\) & \(0.195\pm0.03\) & SMC, intrinsic-slope systematic \\
\({\rm FWHM}_{\rm MgII}\) & \(2684\pm70~\kms\) & absorption-masked fit \\
\(S_\nu(249.276~{\rm GHz})\) & \(2.62\pm0.42~{\rm mJy}\) & continuum aperture \\
\(S_{\rm \cii}\Delta v\) & \(6.47\pm0.23~{\rm Jy\,km\,s^{-1}}\) & single Gaussian fit \\
\({\rm FWHM}_{\rm \cii}\) & \(363.1\pm9.6~\kms\) & single Gaussian fit \\
\cii\ FWHM size & \(0.69\times0.41\) arcsec & deconvolved, \(3.83\times2.27\) kpc \\
\midrule
\multicolumn{3}{l}{\emph{Derived quantities}} \\
\(\lambda L_\lambda(3000\,\angstrom)\) & \(2.14\times10^{46}~{\rm erg\,s^{-1}}\) & dereddened power-law continuum \\
\(\log(M_{\rm BH}/\msun)\) & \(9.04\pm0.42\) & total single-epoch uncertainty \\
\(L_{\rm bol}\) & \(1.10\times10^{47}~{\rm erg\,s^{-1}}\) & \({\rm BC}_{3000}=5.15\) \\
\(\lambda_{\rm Edd}\) & \(0.8\) &  \\
\(L_{\rm \cii}\) & \((7.21\pm0.25)\times10^9~\lsun\) & \\
\(L_{\rm FIR}\) & \((6.3\pm1.0)\times10^{12}~\lsun\) & fiducial dust model, CMB corrected \\
\(L_{\rm TIR}\) & \((8.9\pm1.4)\times10^{12}~\lsun\) & same dust model \\
\(M_{\rm dust}\) & \((4.02\pm0.65)\times10^8~\msun\) & same dust model \\
\({\rm SFR}_{\rm IR}\) & \((1.33\pm0.21)\times10^3~\msun\,{\rm yr^{-1}}\) & \citet{Kennicutt2012} \\
\(M_{\rm dyn,disk}\) & \(2.9\times10^{10}~\msun\) & \(i=60.0^\circ\) \\
\(M_{\rm dyn,isotropic}\) & \(1.1\times10^{11}~\msun\) & dispersion prescription \\
\bottomrule
\end{tabular}
\tablefoot{ALMA uncertainties are approximate \(1\sigma\) statistical errors and exclude absolute calibration and measurement-systematic uncertainties. The line-flux error neglects covariance between the fitted Gaussian amplitude and width. Infrared quantities assume an optically thin modified blackbody with \(T_{\rm dust}(z=0)=47\) K, \(\beta=1.6\), and \(\kappa_{850\mu{\rm m}}=0.77~{\rm cm^2\,g^{-1}}\), including CMB heating and background corrections. Their errors, including that on \({\rm SFR}_{\rm IR}\), propagate only the continuum-flux uncertainty and exclude dust-model and SFR-calibration uncertainties. The nominal SFR is an upper limit because AGN heating is not separated. Both dynamical estimates adopt \(R_{\rm \cii}=0.75\,a_{\rm maj}=2.87\) kpc and the prescriptions in Sect.~\ref{sec:fir}. Size, inclination, and dynamical-mass uncertainties have not been quantified.}
\end{table*}

\section{Discussion}
\label{sec:discussion}


\subsection{The host galaxy of J1440+0019}
\label{sec:fir}

\begin{figure*}
  \centering
  \includegraphics[width=\textwidth]{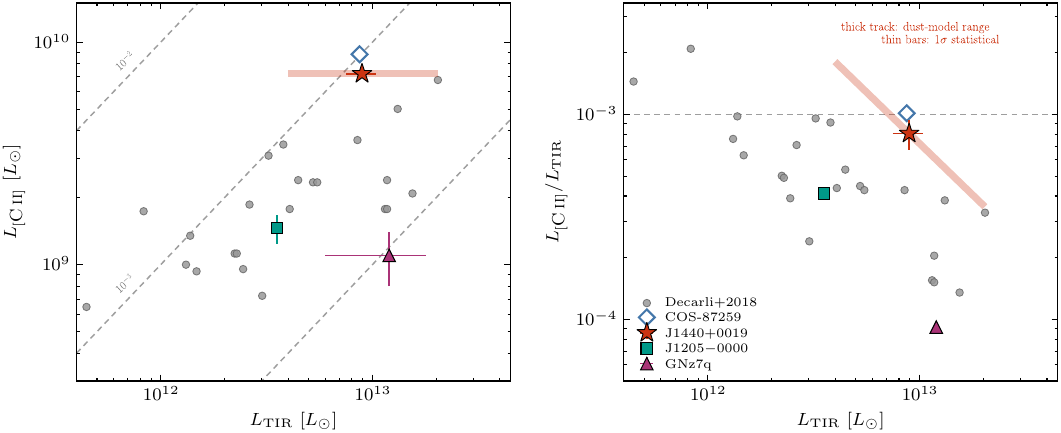}
    \caption{\cii\ and infrared properties. The left panel shows \(\lcii\) against \(\ltir\). Dashed lines mark constant \(\lcii/\ltir\). The right panel shows the same ratio directly. Gray circles are unobscured \(z\simeq6\) quasars from \citet{Decarli2018}, converted from FIR to TIR with the fiducial dust SED. The blue diamond marks COS-87259 \citep{Endsley2023}, whose infrared luminosity corresponds to the star-forming component of its fitted SED. The teal square marks J1205\(-\)0000 \citep{Sawamura2025}, and the purple triangle marks GNz7q using the infrared luminosity from \citet{Fujimoto2022}. The reason for retaining this value rather than the lower estimate from the JWST-based SED fit of \citet{Fei2026} is given in Table~\ref{tab:highz_red_comparison}. The red star marks \yssl. The thick red track shows the variation for \(T_{\rm dust}=35\)--60 K at \(\beta=1.6\), together with additional models at \(\beta=1.5\) and 2.0 for \(T_{\rm dust}=47\) K.}
  \label{fig:lcii_ltir}
\end{figure*}

The ALMA measurements presented in Sect.~\ref{sec:cii} reveal a host galaxy that is luminous in both \cii\ and dust continuum emission. With \(\lcii\simeq 7\times10^9~\lsun\) and a fiducial \(\ltir\simeq 8.9\times10^{12}~\lsun\), \yssl\ lies at the luminous end of the UV-selected \(z\gtrsim6\) quasar sample of \citet{Decarli2018}, as shown in the left panel of Fig.~\ref{fig:lcii_ltir}. We use TIR luminosities integrated over rest-frame 8--1000 \(\mu{\rm m}\) throughout this comparison, converting the published FIR luminosities of \citet{Decarli2018} (42.5--122.5 \(\mu{\rm m}\)) using \(\ltir=1.41\,\lfir\), as appropriate for the same optically thin modified blackbody adopted for \yssl\ with \(T_{\rm dust}=47\) K and \(\beta=1.6\). Figure~\ref{fig:lcii_ltir} also includes several other reddened or obscured high-redshift sources, whose similarities to and differences from \yssl\ are discussed in Sect.~\ref{sec:landscape}.

We next compare the \cii\ and infrared luminosities through the ratio \(\lcii/\ltir\). For the fiducial dust SED, \(\lcii/\ltir\simeq 8.1\times10^{-4}\), which places \yssl\ at the high-ratio end of the \citet{Decarli2018} quasar-host sample at comparable \(\ltir\) (Fig.~\ref{fig:lcii_ltir}, right panel). High-redshift quasar hosts span more than an order of magnitude in this ratio \citep{Decarli2018}, extending down to the low values that define the so-called \cii\ deficit, the suppression of \cii\ relative to the dust continuum seen in the most infrared-luminous galaxies \citep[e.g.,][]{ DiazSantos2013}. As in local luminous infrared galaxies \citep{DiazSantos2017}, the \cii\ equivalent width in the \citet{Decarli2018} sample decreases with increasing far-infrared surface brightness, indicating that the spread in \(\lcii/\ltir\) reflects the compactness of the dust emission and the associated dust temperature. \yssl\ shows no such deficit at face value: its \cii\ luminosity is high for its infrared luminosity relative to quasar hosts of similar \(\ltir\). This placement is insensitive to the adopted dust SED. Over the tested range \(T_{\rm d}=35\)--60\,K, \(\lcii/\ltir\) varies from \(2\times10^{-3}\) to \(4\times10^{-4}\), but with \(\lcii\) fixed a change in \(T_{\rm d}\) moves \yssl\ along the thick red track of slope \(-1\) in the right panel of Fig.~\ref{fig:lcii_ltir}, and it remains in the upper part of the comparison distribution at each corresponding \(\ltir\).

As mentioned in Sect.~\ref{sec:cii}, the continuum emission implies a substantial dust reservoir, with a fiducial mass of approximately \(4\times10^8~\msun\). If the infrared emission is powered entirely by the star formation, the corresponding $\rm SFR_{IR}$ is of order \(10^3~\msun\,{\rm yr^{-1}}\). The contribution of AGN heating could potentially bias this estimate. However, \citet{Yue2026DustHeating} recently showed with high-resolution ALMA observations of the gravitationally lensed quasar J0439+1634 at \(z=6.52\) that AGN heating dominates the dust continuum within approximately 100 pc, while stellar heating dominates at larger radii, so that the AGN contributes only about 13 percent of the integrated continuum at rest-frame \(162.7~\mu{\rm m}\).

Regarding the kinematics, at first glance, the moment 1 map is consistent with a rotating disk with a monotonic gradient of about \(\pm65~\kms\) along \({\rm PA}\simeq155^\circ\), and the position--velocity diagram along this axis shows the same trend (Fig.~\ref{fig:alma}d--f). However, the bright emission spans approximately one synthesized beam, so beam smearing can suppress a true gradient or produce an apparent one by blending spatially distinct components, and the measured amplitude cannot be converted into a rotation velocity. Moreover, the resolved velocity span of \(130~\kms\) is a factor of three smaller than the line FWHM, so most of the line width is not accounted for by the gradient. A dispersion-dominated system, in which the line width reflects random motions rather than rotation, or a merger of two components blended within the beam would produce the same observables. Both scenarios occur among \(z>6\) quasar hosts observed at higher resolution \citep{Neeleman2021}. We therefore do not commit to a single dynamical configuration and quote dynamical masses for both the disk and the dispersion-dominated cases below.


If the host galaxy is a rotating disk, we follow \citet[Eq.~8]{Wang2024}, which adopts the empirical calibration of \citet{Neeleman2021} and assumes a circular velocity \(v_{\rm circ}=0.5\,{\rm FWHM}_{\rm \cii}/\sin i\) and a disk radius \(R_{\rm \cii}=0.75\,a_{\rm maj}\), where \(a_{\rm maj}\) is the deconvolved major-axis FWHM:
\begin{equation}
\begin{aligned}
M_{\rm dyn} &= 5.8\times10^4\left(\frac{{\rm FWHM}_{\rm \cii}}{\kms\,\sin i}\right)^2\left(\frac{R_{\rm \cii}}{\rm kpc}\right)~\msun \\
&\simeq \frac{2.2\times10^{10}}{\sin^2 i}~\msun .
\end{aligned}
\end{equation}
Because the source is only marginally resolved and the deconvolved minor axis is poorly constrained, the axis ratio does not provide a reliable inclination. We therefore adopt a fiducial inclination of \(i=60^\circ\), corresponding to the median for randomly oriented disks, which gives \(M_{\rm dyn}\simeq2.9\times10^{10}~\msun\). At fixed line width and radius, varying the inclination over \(30^\circ<i<70^\circ\) gives \(M_{\rm dyn}\simeq(2.5\)--\(8.8)\times10^{10}~\msun\). This range illustrates the inclination dependence rather than a confidence interval, and smaller inclinations would imply larger masses.

If the host is instead dispersion dominated, we use the isotropic virial estimator \(M_{\rm dyn}=6.7\,\sigma^2R_{\rm \cii}/G\) with \(\sigma={\rm FWHM}_{\rm \cii}/2.355\) and the same \(R_{\rm \cii}\), which gives \(M_{\rm dyn}\simeq1.1\times10^{11}~\msun\). Both estimates remain dependent on the uncertain source size and the assumed dynamical configuration.

\subsection{Accretion properties and the black-hole--host mass relation}
\label{sec:bhhost}

\begin{figure*}
  \centering
  \includegraphics[width=\textwidth]{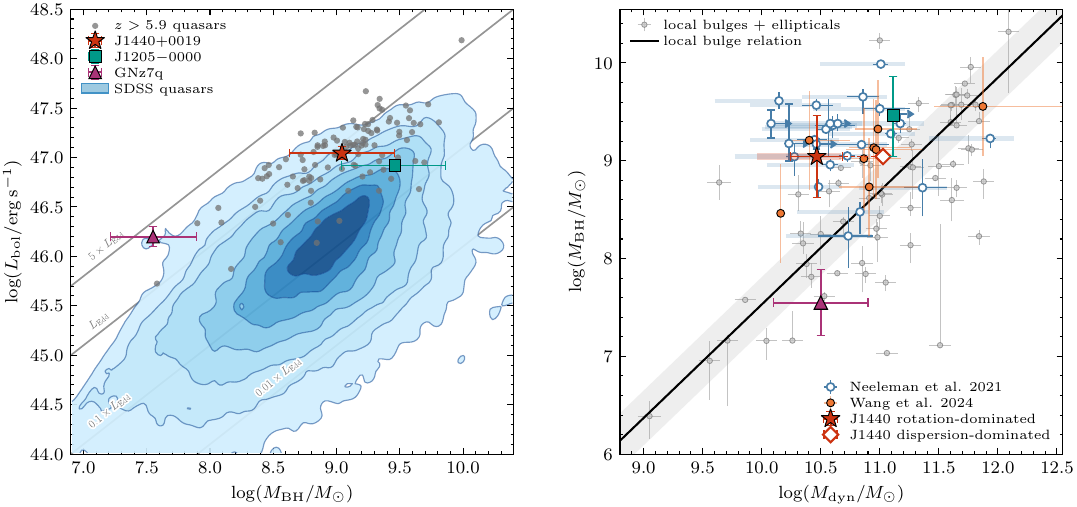}
  \caption{\yssl's SMBH properties. \emph{Left}: Black-hole mass ($M_{\rm BH}$) vs. bolometric luminosity ($L_{\rm bol}$). Contours show the SDSS quasar distribution \citep{Wu2022}, gray points the \(z>5.9\) compilation of \citet{Fan2023}, and gray lines constant Eddington ratios. \emph{Right}: black-hole mass against \cii\ dynamical mass for the \(z\gtrsim6\) quasars of \citet{Neeleman2021} (blue, with arrows marking lower limits) and \citet{Wang2024} (orange). The black line and shaded band show the local \(M_{\rm BH}\)--\(M_{\rm bulge}\) relation of \citet{Kormendy2013} and its 0.3 dex scatter, with local galaxies shown in gray. \yssl\ is marked by a red star. In the right panel, its disk and isotropic dynamical-mass estimates are shown by the filled star and open diamond, respectively, joined by a dashed line. The horizontal bar shows the adopted range of the disk mass estimate, and the vertical bar represents the total black-hole-mass uncertainty. The teal square marks J1205\(-\)0000 using its extinction-corrected nuclear properties \citep{Kato2020} and, in the right panel, the disk dynamical mass from \citet{Sawamura2025}. The purple triangle marks GNz7q \citep{Fei2026}, whose host stellar mass is used in the right panel. The local relation is shown for reference, since the high-redshift host masses do not measure stellar bulge masses directly.}
  \label{fig:mbh_mdyn}\label{fig:mbh_lbol}
\end{figure*}

The spectral modeling in Sect.~\ref{sec:mbh} yields an extinction-corrected black-hole mass of \(\log(M_{\rm BH}/\msun)=9.04\pm0.42\), a bolometric luminosity of \(L_{\rm bol}\simeq1.1\times10^{47}~{\rm erg\,s^{-1}}\), and a corresponding Eddington ratio of \(\lambda_{\rm Edd}\simeq0.8\). The left panel of Fig.~\ref{fig:mbh_lbol} places these values among the \(z>5.9\) quasars using the compilation from \citet{Fan2023}, with the SDSS quasar distribution as a lower-redshift reference \citep{Wu2022}. The two comparison sources J1205\(-\)0000 and GNz7q are discussed in Sect.~\ref{sec:landscape}. \yssl\ lies within the distribution of high-redshift blue quasars (gray points), with an Eddington ratio comparable to those of blue quasars at similar redshift and luminosity.

The right panel of Fig.~\ref{fig:mbh_mdyn} shows \(M_{\rm BH}\) against the two host dynamical-mass estimates derived in Sect.~\ref{sec:fir}, \(2.9\times10^{10}~\msun\) for a rotating disk at the fiducial inclination of \(i=60^\circ\) and \(1.1\times10^{11}~\msun\) for a dispersion-dominated system. Measurements of blue \(z\gtrsim6\) quasars from \citet{Neeleman2021} (blue) and \citet{Wang2024} (orange) are shown for comparison, together with the local relation. For luminous UV-selected \(z\gtrsim6\) quasars, the measured black-hole and \cii\ dynamical masses place the population above the local relation by about an order of magnitude on average, with a spread of about two orders of magnitude (e.g., \citealt{Neeleman2021,Wang2024}; but also see \citealt{Lauer2007,Izumi2021b,Silverman2025}).

For \yssl, both estimates nominally place this quasar in the overmassive regime (although the dispersion-dominated estimate lies within the intrinsic scatter of the local relation). That also means that in both panels of Fig.~\ref{fig:mbh_mdyn} this red quasar is indistinguishable from its blue counterparts. These conclusions rest on a single object. If the reddening marks a phase in which the black hole grows ahead of its host, red quasars as a class should sit at higher Eddington ratios or higher \(M_{\rm BH}/M_{\rm dyn}\) than blue quasars of the same luminosity. Testing this requires a sample of \(z>6\) red quasars with black-hole and host masses measured in the same way and placed on the same two planes; at present J1205\(-\)0000 and GNz7q are the only other entries (Sect.~\ref{sec:landscape}).

\subsection{Connection to lower-redshift red quasars}
\label{sec:dustloc}\label{sec:lowz}

The nuclear reddening of \yssl, with \(E(B-V)\simeq0.20\) and \(A_V\simeq0.54\) for the adopted SMC extinction law, lies within the range \(A_V\simeq0.06\)--0.7 measured for optically selected red quasars at \(z\simeq1.5\) by \citet{Fawcett2022}. There is not yet a consensus on whether lower-redshift red quasars differ systematically from their blue counterparts in their accretion properties. While \citet{Fawcett2022} find no evidence for enhanced accretion in their moderately reddened X-shooter sample, \citet{Kim2024} infer higher Eddington ratios for FIRST--2MASS red quasars at \(z\lesssim1\) than for unobscured quasars matched in luminosity and redshift, using estimators based on the mid-infrared continuum. The Eddington ratio of \yssl, \(\lambda_{\rm Edd}\simeq0.8\), exceeds the values estimated for most red quasars in these lower-redshift samples, but remains consistent with those of luminous blue quasars at similar redshift (Sect.~\ref{sec:bhhost}). 

The enhanced incidence of compact radio emission in lower-redshift red quasars has also been interpreted as evidence for intrinsic differences from blue quasars \citep{Klindt2019,Fawcett2020}. \yssl\ is undetected in the available radio surveys, although the upper limits are too shallow to exclude comparable radio emission (Appendix~\ref{app:radio}). Broader conclusions about whether high-redshift red quasars differ systematically from their blue counterparts remain premature given the small number of objects studied so far and the limited radio constraints available for \yssl.

As for host galaxy properties, \citet{CalistroRivera2021} find no clear differences in star formation or the interstellar medium between moderately reddened quasars at lower redshift and blue controls matched in luminosity and redshift, based on broad-band SED modelling. Spatially resolved millimetre and submillimetre observations of matched samples would further test whether moderately reddened and blue quasars have similar cold-gas reservoirs and dust distributions. A complementary perspective comes from the more heavily reddened quasars. \citet{Banerji2017} detected dust continuum and CO emission in four quasars at \(z\simeq2.5\) with \(A_V\simeq2.5\)--6, finding a mean dust mass of \(6\times10^8~\msun\) assuming the same dust SED adopted here (\(T_{\rm dust}=47\) K and \(\beta=1.6\)). The fiducial dust mass of \yssl, \(4\times10^8~\msun\), is comparable, despite its much weaker nuclear extinction. The inferred star formation rates of those quasars are \(\gtrsim10^3~\msun\,{\rm yr^{-1}}\), while the infrared luminosity of \yssl\ implies a rate of the same order, with \({\rm SFR_{IR}}\simeq1330~\msun\,{\rm yr^{-1}}\). 

Finally, the high-ionization absorption systems in \yssl, reaching blueshifts of \(\simeq3600~\kms\), suggest a further connection through outflows. At lower redshift, the rest-frame 2--5~\(\mu{\rm m}\) excess in red quasars is strongest among objects with wind signatures, motivating a contribution from dusty winds to their reddening \citep{CalistroRivera2021}. Recent measurements also link faster [\ion{O}{iii}] outflows to stronger infrared excess, while finding no significant correlation with \(E(B-V)\) alone \citep{Fawcett2026}. Evidence that such activity can affect the gas supply comes from \citet{Stacey2022}, who identify compact outflows of warm molecular gas with velocities of 500--1000~\(\kms\) in red quasars at \(z\sim2.5\). In \yssl, the blueshifted absorption supports an outflow interpretation, but neither the location nor the dust content of the absorbing gas is constrained. The current infrared constraints are also insufficient to establish whether \yssl\ exhibits a comparable rest-frame 2--5~\(\mu{\rm m}\) excess (Fig.~\ref{fig:sed_comparison}). No broad \cii\ component is detected within the observed velocity range, and the ALMA spectrum does not cover the velocities of the UV absorption systems (Sect.~\ref{sec:cii}). A dusty wind could potentially contribute to the reddening, but the present data do not establish its role in obscuring the nucleus or clearing the host.

\subsection{Reddened and obscured quasars at \(z>6\)}
\label{sec:landscape}

\begin{table*}[t]
\caption{Nuclear and host properties of selected reddened or obscured sources at \(z>6\). Measurements for \yssl\ are from this work. Comparison data are taken from \citet{Kato2020,Sawamura2025} for J1205$-$0000, \citet{Fujimoto2022,Fei2026} for GNz7q, \citet{Endsley2023} for COS-87259, and \citet{Matsuoka2025} for the obscured SHELLQs sources.}
\label{tab:highz_red_comparison}
\centering
\begingroup
\small
\setlength{\tabcolsep}{4pt}
\renewcommand{\arraystretch}{1.22}

\begin{tabularx}{\textwidth}
{@{}p{0.21\textwidth}*{5}{>{\centering\arraybackslash}X}@{}}
\toprule
Target(s)
& J1440\(+\)0019
& J1205\(-\)0000
& GNz7q
& COS-87259
& SHELLQs tpye-II candidates\tablefootmark{a} \\
\midrule

Redshift
& \(6.5488\pm0.0001\)
& 6.7224
& 7.1899
& 6.8532
& 6.075--6.875 \\

\addlinespace[2pt]
\multicolumn{6}{@{}l}{\emph{Nucleus}} \\

Observed AGN signature
& UV broad lines
& UV broad lines; BALs
& Balmer broad lines
& IR/radio AGN
& Balmer/He\,I broad lines \\

Continuum \(E(B-V)\)
& \(0.195\pm0.030\)
& \(0.115^{+0.029}_{-0.027}\)
& \(\ldots\)
& \(\ldots\)
& \(\ldots\) \\

Extinction indicator\tablefootmark{b}
& \(A_V=0.53\pm0.08\)
& \(A_V=0.34\pm0.08\)
& \(A_V=0.21^{+0.09}_{-0.08}\)
& \(\tau_{9.7}=2.3\pm0.1\)
& \(A_V=1.0\text{--}2.2\) \\

\(\log(M_{\rm BH}/M_\odot)\)\tablefootmark{c}
& \(9.04\pm0.42\)
& \(9.46^{+0.40}_{-0.42}\)
& \(7.55\pm0.34\)
& \(\ldots\)\tablefootmark{d}
& 7.80--9.11 \\

\(\lambda_{\rm Edd}\)\tablefootmark{c}
& \(0.8^{+1.3}_{-0.5}\)
& \(0.22^{+0.34}_{-0.13}\)
& \(2.7\pm0.4\)
& \(\ldots\)\tablefootmark{d}
& 0.33--1.13 \\

\addlinespace[3pt]
\multicolumn{6}{@{}l}{\emph{Host dust and gas}} \\

\(L_{\rm TIR}\;[10^{12}\,L_\odot]\)\tablefootmark{e}
& \(8.9\pm1.4\)
& \(3.54\pm0.05\)
& \(12\pm6\)\tablefootmark{f}
& \(9.1\pm1.1\)
& \(\ldots\) \\

\(L_{\rm [CII]}\;[10^9\,L_\odot]\)
& \(7.21\pm0.25\)
& \(1.46\pm0.22\)
& \(1.1\pm0.3\)
& \(8.82\pm0.75\)
& \(\ldots\) \\

\bottomrule
\end{tabularx}
\endgroup

\tablefoot{
\tablefoottext{a}{Ranges refer to the nine broad-line objects G01--G07, Q01, and Q02, excluding the four narrow-line sources.}
\tablefoottext{b}{For J1440 and J1205, \(A_V=R_V E(B-V)\), with \(R_V=2.74\) and 2.93, respectively, following the adopted SMC laws. J1440's reddening error reflects intrinsic continuum-slope uncertainty. GNz7q's \(A_V\) is from narrow Balmer lines; the SHELLQs range is from BLR Balmer decrements. COS-87259's \(\tau_{9.7}\) is an SED-model parameter, consistent with Compton-thick obscuration but not a measurement of \(N_{\rm H}\).}
\tablefoottext{c}{J1440's mass error includes 0.4 dex of single-epoch calibration scatter; its Eddington-ratio interval propagates the total 0.42 dex mass error at fixed bolometric luminosity. For J1205, we add 0.4 dex in quadrature to the published mass and Eddington-ratio errors in logarithmic space, consistent with the systematic uncertainty represented in Fig.~\ref{fig:mbh_lbol}. These intervals exclude bolometric-correction systematics. GNz7q retains the published errors and uses an accretion-rate-dependent H\(\beta\) calibration with \(L_{\rm Edd}/M_{\rm BH}=1.5\times10^{38}\,{\rm erg\,s^{-1}}\,M_\odot^{-1}\).}
\tablefoottext{d}{COS-87259 has no virial mass measurement. Its SED instead implies \(M_{\rm BH}\lambda_{\rm Edd}\simeq(1.6\pm0.2)\times10^9\,M_\odot\) under the adopted bolometric model.}
\tablefoottext{e}{All entries correspond to rest-frame 8--1000\,\(\mu{\rm m}\). J1440 and J1205 use modified blackbodies with \(T_{\rm dust}=47\) K and \(\beta=1.6\), including CMB corrections and no subtraction for AGN heating. COS-87259 uses the star-forming component of its decomposed SED; GNz7q includes its fitted AGN component.}
\tablefoottext{f}{GNz7q's adopted total luminosity is from the original optical-to-mm fit; its star-forming component is \((11\pm5)\times10^{12}\,L_\odot\). The newer JWST-based energy-balance fit yields \(\log(L_{\rm FIR}/L_\odot)=12.3\pm0.3\), but underpredicts the 850\,\(\mu{\rm m}\) flux and does not specify integration limits, so it is not substituted here.}
}
\end{table*}

\begin{figure*}[htp!]
  \centering
  \includegraphics[width=\textwidth]{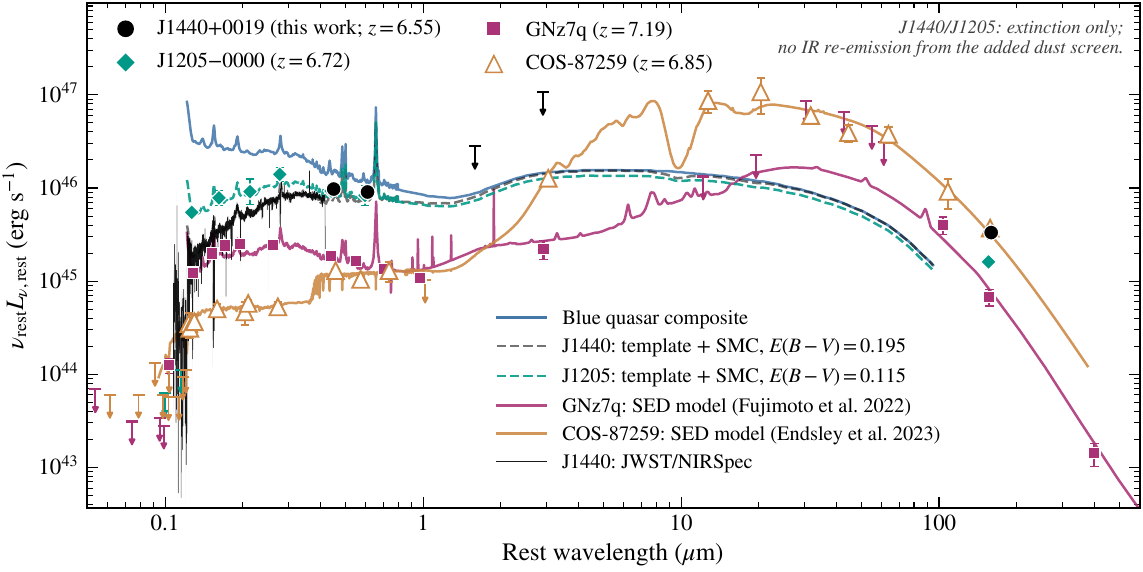}
\caption{Rest-frame UV-to-far-infrared spectral energy distribution models of \yssl\ (black), J1205$-$0000 (teal), GNz7q (magenta), and COS-87259 (orange). Symbols represent photometric measurements, with arrows indicating 2$\sigma$ upper limits. The black curve shows the JWST/NIRSpec spectrum of \yssl. The blue curve shows an blue quasar template \citep{VandenBerk2001,Richards2006,Lusso2015}, while the gray and teal dashed curves show templates attenuated by SMC-like extinction with \(E(B-V)=0.195\) and 0.115 for \yssl\ and J1205$-$0000, respectively. These dashed curves do not include additional infrared emission from the dust responsible for the reddening and should not be interpreted as complete SED models. The magenta and orange curves show the published SED models for GNz7q \citep{Fujimoto2022} and COS-87259 \citep{Endsley2023}.}
\label{fig:sed_comparison}
\end{figure*}

Table~\ref{tab:highz_red_comparison} compares \yssl\ with other reddened and obscured sources at \(z>6\). Figure~\ref{fig:sed_comparison} compares the rest-frame UV-to-infrared spectral energy distributions of \yssl, J1205$-$0000, GNz7q, and COS-87259, illustrating the differences in their observed nuclear and dust emission. The closest match is J1205$-$0000 at \(z=6.72\), whose continuum is likewise described by SMC-like reddening, with \(E(B-V)=0.115\) \citep{Kato2020}. Both show blueshifted high-ionization absorption, although J1205$-$0000 is a BAL quasar \citep{Onoue2019}, whereas \yssl\ exhibits narrower, mini-BAL absorption. J1205$-$0000 has a nominally larger black-hole mass and a lower Eddington ratio, \(\lambda_{\rm Edd}\simeq0.22\), although the uncertainties are substantial (Fig.~\ref{fig:mbh_lbol}). The hosts also differ: the ALMA measurements imply that \yssl\ is approximately 2.5 times more infrared-luminous and 4.9 times more \cii-luminous \citep{Sawamura2025} (see Fig.~\ref{fig:lcii_ltir}). Similar nuclear reddening can therefore accompany different inferred accretion properties and host luminosities. Neither object has an established fast, galaxy-scale \cii\ outflow in the existing data, so their shared nuclear absorption does not yet demonstrate feedback on the cold gas of their hosts.

GNz7q at \(z=7.19\) illustrates a different combination of black-hole and host growth. Its inferred black-hole mass, \(\log(M_{\rm BH}/\msun)=7.55\pm0.34\), is about 30 times smaller than that of \yssl, while its nominal Eddington ratio is higher, \(\lambda_{\rm Edd}\simeq2.7\) (Fig.~\ref{fig:mbh_lbol}). Its host has \(\log(M_\star/\msun)=10.5\pm0.4\), giving \(M_{\rm BH}/M_\star\simeq10^{-3}\) \citep{Fei2026}. Its nominal position lies below the local bulge relation plotted in Fig.~\ref{fig:mbh_lbol}, whereas the nominal positions of \yssl\ and J1205$-$0000 lie above it; however, this comparison uses stellar mass for GNz7q and dynamical masses for the other two sources. GNz7q is also markedly X-ray weak \citep{Fujimoto2022}, and its UV--optical continuum is not well reproduced by standard quasar templates with conventional extinction curves \citep{Fei2026}. These differences show that observed red colors do not by themselves imply a common extinction law or accretion state at high redshift.

COS-87259 at \(z=6.85\) provides a complementary comparison through its host emission. Its star-forming infrared component, \(L_{\rm TIR}=(9.1\pm1.1)\times10^{12}~\lsun\), and \cii\ luminosity, \((8.82\pm0.75)\times10^9~\lsun\), are comparable to those of \yssl\ \citep{Endsley2023} (Fig.~\ref{fig:lcii_ltir}), although the infrared estimate for \yssl\ may include AGN heating. Its nucleus, however, is heavily obscured and identified through its infrared and radio emission \citep{Endsley2022,Endsley2023,Varnava2024}. This contrast is also evident in Fig.~\ref{fig:sed_comparison}. While the UV--optical continuum of \yssl\ resembles that of J1205$-$0000, its monochromatic luminosity near rest-frame \(160~\mu{\rm m}\) is comparable to that of COS-87259, whose nucleus is much more heavily obscured. 

\citet{Matsuoka2025} present SHELLQs sources at \(z>6\) with strong Ly\(\alpha\) emission and very weak UV--optical continua relative to their emission lines. JWST spectroscopy reveals broad Balmer components in nine objects, while four exhibit only narrow lines. Several broad-line sources resemble type~1.8 quasars; the narrow-line objects remain type-II candidates. The broad-line objects have inferred black-hole masses of \(\log(M_{\rm BH}/\msun)=7.80\)--9.11 and Eddington ratios of \(\lambda_{\rm Edd}=0.33\)--1.13, overlapping the properties of \yssl. Their Balmer decrements imply BLR extinctions of \(A_V\simeq1.0\)--2.2 mag, and Chandra observations yield non-detections in four objects \citep{Iwasawa2026}. These observations motivate an interpretation in which the continuum source and part of the BLR are hidden, although the nature and location of the obscuring material remain uncertain. ALMA observations could constrain the host dust reservoir and gas distribution; combined with measurements of hot-dust emission, they would help assess whether obscuration is concentrated around the nucleus or extends across the host.

The infrared SED of \yssl\ remains poorly constrained between the available near-infrared photometry and the ALMA continuum measurement (Fig.~\ref{fig:sed_comparison}). The reddened templates shown for \yssl\ and J1205$-$0000 account for extinction along the nuclear sightline but do not include additional thermal emission from the dust responsible for that extinction. Neither the luminosity nor the spectral shape of this component can be inferred from \(E(B-V)\) alone, because they also depend on the dust covering factor, temperature distribution, and geometry. Better sampling of the infrared SED would help separate hot and warm AGN-heated dust from colder host emission and constrain the possible origin of the reddening.

The strong detections of both \cii\ and dust continuum emission in \yssl\ with only 5.5~min of ALMA on-source integration demonstrate that shallow ALMA surveys of known high-redshift quasars can efficiently identify similarly dust-rich systems. Such surveys can help flag promising reddened-quasar candidates for spectroscopic follow-up at a modest cost in ALMA observing time. Euclid is capable of building up samples of mildly reddened quasars like \yssl, while Roman's greater depth can extend these searches to more heavily obscured sources. The discoveries of GNz7q and COS-87259 already suggest that a substantial population may be missed by conventional UV-selected quasar surveys \citep{Fujimoto2022,Endsley2023}. JWST/MIRI searches, including SMILES and MEOW, are revealing additional high-redshift obscured AGN candidates \citep{Lyu2024,Leung2026}, with MEOW luminosity-function estimates implying obscured fractions as high as 98--99\% at \(4.5<z<6\) \citep{Bulichi2026}. Rest-frame UV and optical spectroscopy remains essential to establish the nature of these candidates and constrain their extinction and accretion properties. The Euclid and Roman deep fields can bridge the gap in area and depth between wide surveys and JWST pencil-beam fields. Combined with spectroscopic and longer-wavelength follow-up, they could help build a more complete census and establish how much early black-hole growth occurs in obscured systems.

\section{Conclusions}
\label{sec:conclusions}

We have combined JWST/NIRSpec spectroscopy and ALMA observations to characterize the nucleus and host of the reddened quasar \yssl\ at \(z=6.5488\). Our main results are:

\begin{enumerate}
\item The continuum is best described by SMC-like extinction with \(E(B-V)=0.195\pm0.03\), where the uncertainty reflects the unknown intrinsic continuum slope. The spectrum shows no evidence for a 2175~\(\angstrom\) extinction feature.

\item Three high-ionization absorption systems exhibit blueshifts reaching approximately \(3600~\kms\). Their \ion{N}{v}, \ion{Si}{iv}, and \ion{C}{iv} absorption supports an intrinsic or associated origin, although the location and dust content of the absorbing gas remain unconstrained.

\item The extinction-corrected \ion{Mg}{ii} fit gives \(\log(M_{\rm BH}/\msun)=9.04\pm0.42\). Applying a bolometric correction of \({\rm BC}_{3000}=5.15\) to the dereddened continuum luminosity at 3000 \(\angstrom\) gives \(L_{\rm bol}=1.1\times10^{47}~{\rm erg\,s^{-1}}\), corresponding to an Eddington ratio of \(\lambda_{\rm Edd}\simeq0.8\).

\item The host has \(\lcii=7.21\times10^9~\lsun\). For \(T_{\rm dust}=47\) K and \(\beta=1.6\), its continuum implies \(\ltir=8.9\times10^{12}~\lsun\), \(M_{\rm dust}\simeq 4.0\times10^8~\msun\), and an upper limit of approximately \(1330~\msun\,{\rm yr^{-1}}\) on the SFR. The \cii\ emission is marginally resolved in space, with disk and dispersion-dominated prescriptions giving \(M_{\rm dyn}\simeq 2.9\times10^{10}\) and \(1.1\times10^{11}~\msun\), respectively. No broad \cii\ component is detected within the observed velocity range, which does not cover the UV absorption systems.

\item The Eddington ratio and position of \yssl\ in the \(M_{\rm BH}\)--\(M_{\rm dyn}\) plane are similar to those of high-redshift blue quasars. Its host lies at the luminous end of their distribution, with a relatively high \(\lcii/\ltir=8.1\times10^{-4}\) at comparable \(\ltir\). The dispersion-dominated mass estimate remains consistent with the intrinsic scatter of the local black-hole--bulge relation.

\item The nuclear reddening of \yssl\ resembles that of moderately reddened low-redshift quasars, while its dust reservoir is comparable to those of more heavily reddened systems. Its nominal Eddington ratio exceeds those of most objects in the low-redshift comparison samples. The radio limits are too shallow to test the compact radio excess reported for low-redshift red quasars.

\item Among reddened and obscured sources at \(z>6\), \yssl\ resembles J1205$-$0000 in its nuclear spectrum but has a more luminous host, comparable to COS-87259. GNz7q instead has a smaller black hole and a higher nominal Eddington ratio, while the obscured SHELLQs quasars have much weaker UV continua relative to their emission lines. These comparisons show that red and obscured sources span diverse accretion states and host properties.
\end{enumerate}

\begin{acknowledgements}

DY and JFH acknowledge support from the European Research Council (ERC) under the European Union's Horizon 2020 research and innovation program (grant agreement No.~885301). JFH acknowledges support from NSF grant No.~2307180.

This work is based on observations made with the NASA/ESA/CSA James Webb Space Telescope, associated with program GO~9180. The data were obtained from the Mikulski Archive for Space Telescopes at the Space Telescope Science Institute, which is operated by the Association of Universities for Research in Astronomy, Inc., under NASA contract NAS~5-03127 for JWST. This paper makes use of the following ALMA data: ADS/JAO.ALMA\#2023.1.00443.S. ALMA is a partnership of ESO (representing its member states), NSF (USA) and NINS (Japan), together with NRC (Canada), NSTC and ASIAA (Taiwan), and KASI (Republic of Korea), in cooperation with the Republic of Chile. The Joint ALMA Observatory is operated by ESO, AUI/NRAO and NAOJ. 

Claude (Anthropic) and ChatGPT (OpenAI) were used to assist with language editing and revision of the manuscript text and with the development of analysis/plotting scripts used to produce the figures. All scientific analysis, results, and conclusions are the authors' own, and the authors take full responsibility for the content of this paper.

\end{acknowledgements}

\section*{Data availability}
The JWST/NIRSpec data underlying this article are available in the Mikulski Archive for Space Telescopes (MAST) under program GO 9180, and the ALMA data are available in the ALMA Science Archive under project code 2023.1.00443.S. The reduced spectra, calibrated data cubes, and other data products generated in this work will be shared on reasonable request to the corresponding author.

\bibliographystyle{bibtex/aa}
\bibliography{J1440_refs}

\begin{appendix}
\onecolumn

\clearpage
\section{Comparison with archival photometry}
\label{app:photometry}

\begin{figure}[H]
    \centering
    \includegraphics[width=\textwidth]{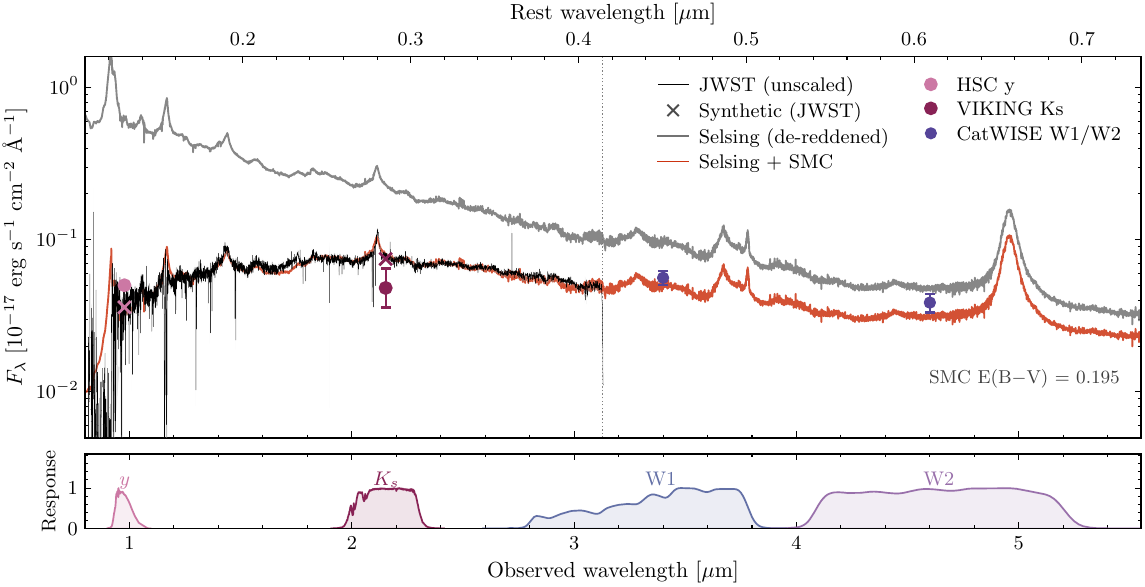}
    \caption{Comparison of the JWST/NIRSpec spectrum with archival photometry. The black curve shows the spectrum before rescaling to the HSC \(y\)-band photometry. Colored circles show the HSC \(y\), VIKING DR5 \(K_s\), and CatWISE2020 \(W1\) and \(W2\) measurements, while crosses show synthetic \(y\)- and \(K_s\)-band photometry derived from the spectrum. The gray and red curves show the unreddened \citet{Selsing2016} composite and the same composite attenuated by an SMC extinction law with \(E(B-V)=0.195\), respectively. The lower panel shows the filter transmission curves.}
    \label{fig:photometry}
\end{figure}

\section{Comparison of extinction laws}
\label{app:extlaws}

We test the dependence of the continuum fit on the adopted extinction law by repeating the procedure in Sect.~\ref{sec:reddening} with five alternative extinction and attenuation curves. These comprise the LMC average and LMC2 supershell curves of \citet{Gordon2003}, with \(R_V=3.41\) and 2.76, respectively, the Milky Way curve of \citet{Fitzpatrick1999} with \(R_V=3.1\), the starburst attenuation law of \citet{Calzetti2000} with \(R_V=4.05\), and the mean AGN curve of \citet{Gaskell2007}. For the last curve, we assume \(R_V=3.1\) when reporting \(A_V\). All fits use the same \citet{Selsing2016} composite, continuum windows, absorption masks, and spectral uncertainties as the fiducial SMC fit. Only \(E(B-V)\) and the normalization are varied, with the intrinsic template shape held fixed.

\begin{figure}[H]
  \centering
  \includegraphics[width=0.82\textwidth]{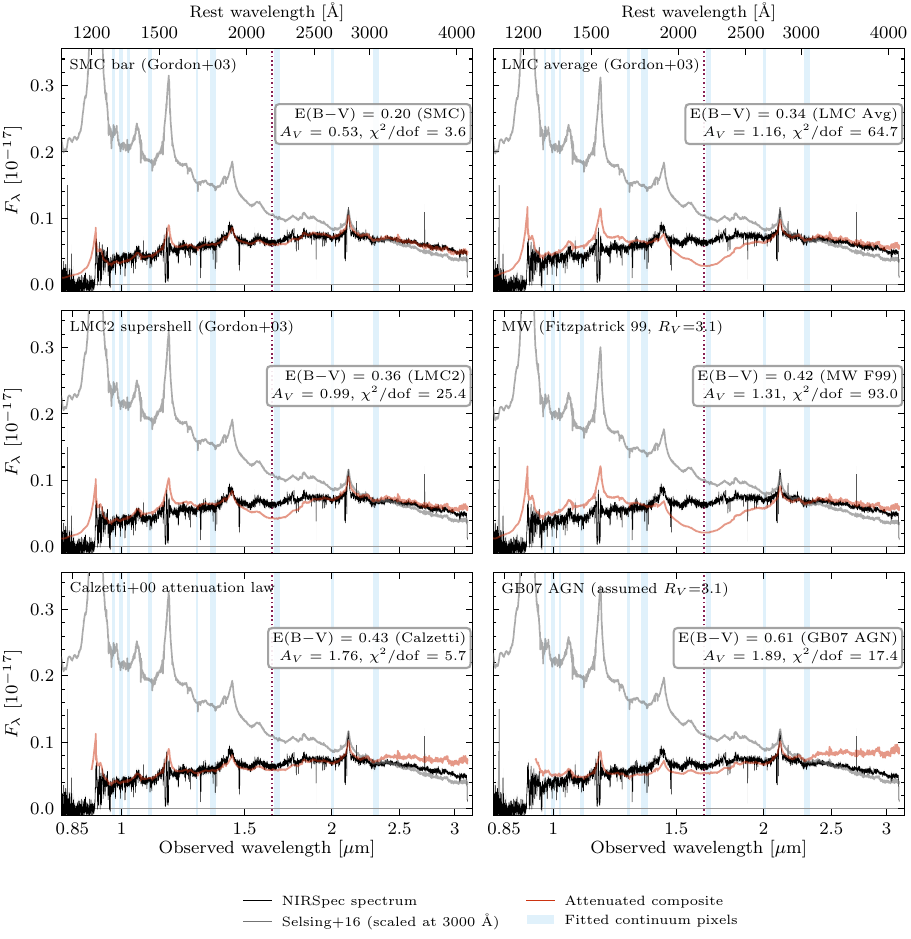}
  \caption{Comparison of the extinction and attenuation laws fitted to the NIRSpec spectrum of \yssl. Black curves show the observed spectrum, red curves the best-fitting attenuated composite, and gray curves the unreddened \citet{Selsing2016} composite scaled to match each attenuated model at rest-frame 3000~\(\angstrom\). Blue shaded regions mark the continuum pixels used in the fits, and the purple dotted line marks the expected position of the 2175~\(\angstrom\) feature. Each panel lists the fitted \(E(B-V)\), the corresponding \(A_V\), and the reduced \(\chi^2\). Flux densities are in units of \(10^{-17}~{\rm erg\,s^{-1}\,cm^{-2}\,\angstrom^{-1}}\).}
  \label{fig:extlaws}
\end{figure}

Figure~\ref{fig:extlaws} compares the resulting fits. The SMC curve gives the lowest reduced \(\chi^2\), with \(\chi^2/\nu=3.6\), where \(\nu\) is the number of fitted pixels minus two. The LMC average, LMC2, and Milky Way curves give \(\chi^2/\nu=64.7\), 25.4, and 93.0, respectively. Their main discrepancy is the broad depression associated with the 2175~\(\angstrom\) extinction feature, which falls at an observed wavelength of \(1.64~\mu{\rm m}\) at the systemic redshift of \yssl. The observed continuum shows no corresponding trough, including at the weaker depth predicted by the LMC2 curve.

The featureless Calzetti and mean AGN curves give \(\chi^2/\nu=5.7\) and 17.4, respectively. Their best-fit color excesses are \(E(B-V)=0.43\) and 0.61, corresponding to \(A_V=1.76\) and 1.89 mag under the adopted \(R_V\) values, compared with \(A_V=0.53\) mag for the SMC fit. Both also overpredict the continuum toward the red end of the displayed spectrum, beyond the fitted windows. The quoted fit statistics include the spectral uncertainties but do not account for intrinsic quasar spectral diversity.

\clearpage

\section{Low-ionization absorption system}
\label{app:lowion}

\begin{figure}[H]
  \centering
  \includegraphics[width=\textwidth]{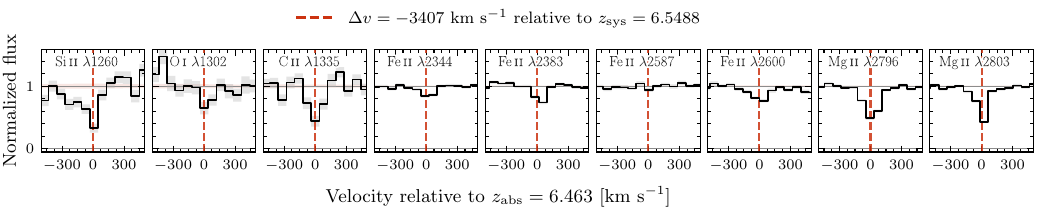}
  \caption{Continuum-normalized absorption profiles around the low-ionization transitions discussed in Sect.~\ref{sec:absorption}. Black stepped curves show the normalized spectrum, and the horizontal lines mark unity. Velocities are measured relative to \(z_{\rm abs}=6.463\), marked by the red dashed vertical lines.}
  \label{fig:lowion}
\end{figure}

Figure~\ref{fig:lowion} shows the continuum-normalized profiles of the low-ionization absorption discussed in Sect.~\ref{sec:absorption}, with velocities measured relative to \(z_{\rm abs}=6.463\). Both members of the \ion{Mg}{ii} doublet and the accompanying metal lines show absorption near this redshift, supporting their identification as a common system.

\section{Radio cutouts}
\label{app:radio}

\begin{figure}[H]
  \centering
  \includegraphics[width=0.48\textwidth]{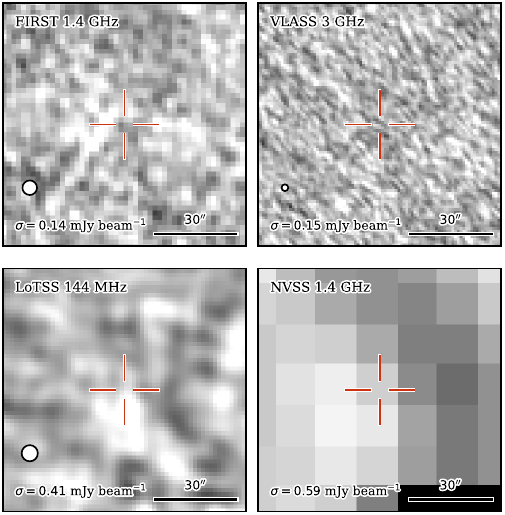}
  \caption{Radio cutouts centered on \yssl\ from FIRST (1.4 GHz), VLASS (3 GHz), LoTSS DR3 (144 MHz), and NVSS (1.4 GHz). Each panel spans \(1.5\times1.5\) arcmin, with the crosshair marking the optical position and a scale bar indicating \(30\) arcsec. The grayscale spans \(-2\sigma\) to \(+5\sigma\), where \(\sigma\) is the local rms noise annotated in each panel. White circles indicate the beam FWHM, except for the \(45\) arcsec NVSS beam, which is omitted. No source is detected at the quasar position in any panel.}
  \label{fig:radio}
\end{figure}

\yssl\ is undetected in FIRST, VLASS, LoTSS DR3, and NVSS (Fig.~\ref{fig:radio}). Forced photometry at the optical position gives a \(3\sigma\) upper limit of \(S_{1.4\,{\rm GHz}}<0.43\) mJy from FIRST for an unresolved source. Assuming a power-law radio spectrum, \(S_\nu\propto\nu^\alpha\), with \(\alpha=-0.29\), this corresponds to a rest-frame luminosity of \(L_{1.4\,{\rm GHz}}<5\times10^{25}~{\rm W\,Hz^{-1}}\).

At lower redshift, red quasars show a higher radio-detection fraction than blue quasars matched in luminosity and redshift, with the excess arising primarily from compact sources near the radio-quiet to radio-loud boundary \citep{Klindt2019,Fawcett2020}. For comparison, the approximately 1 mJy FIRST detection threshold corresponds to \(L_{1.4\,{\rm GHz}}\simeq10^{25}~{\rm W\,Hz^{-1}}\) at \(z\simeq1.5\), several times below our upper limit. The present non-detections therefore do not exclude radio emission comparable to that of the fainter sources contributing to the low-redshift excess.

\end{appendix}

\end{document}